\documentclass{PoS}
\usepackage{graphicx}
\usepackage{amsmath}
\usepackage{amsfonts}
\usepackage{amssymb}
\usepackage{amsthm}
\usepackage{caption}
\usepackage{fontenc}
\usepackage{graphicx}
\usepackage{comment}
\usepackage{wrapfig}
\usepackage{epsfig}

\usepackage[most]{tcolorbox}
\usepackage{textpos}
\usepackage{longfbox} 
\title{Glueballs as gravitons in holographic approaches}

\ShortTitle{Glueballs as gravitons in holographic approaches}

\author{\speaker{M. Rinaldi}\\
        Dipartimento di Fisica e Geologia. Universit\`a degli studi di 
        Perugia. INFN section of Perugia. Via A. Pascoli, Perugia, 
        Iataly.\\
        E-mail: \email{matteo.rinaldi@pg.infn.it}}

\abstract{
In this contribution we present a 
a phenomenological analysis of the scalar glueball and scalar meson 
spectra
within AdS/QCD models 
 in the bottom-up 
approach.
In particular, 
we consider a Light-Front QCD framework which allows to relate the 
AdS/QCD mode functions to the hadronic wave-functions.
Such a procedure is here adopted to analyse the mixing between scalar 
mesons and glueballs. }

\FullConference{Light Cone 2019 - QCD on the light cone: from hadrons to 
heavy ions - LC2019\\
		16-20 September 2019\\
		Ecole Polytechnique, Palaiseau, France}

\begin{document}
\section{Introduction}
In this contribution we investigate the structure of glueballs,  
for which some 
information on
 their spectra and properties have been collected in particular 
within 
the lattice approach ~
\cite{Morningstar:1999rf,Chen:2005mg,Lucini:2004my}. However, 
if glueballs exist they would mix with meson
states with
same quantum numbers
making difficult their direct 
characterization.
Here we consider 
AdS/QCD models 
 to access 
non perturbative features of 
glueballs~\cite{Vento:2017ice,Rinaldi:2017wdn}.
The holographic principle relies in a correspondence between a five
 dimensional 
classical theory with an AdS metric and a supersymmetric conformal 
quantum 
field theory. Since the latter is 
different from 
QCD, we consider a bottom-up 
~\cite{Brodsky:2003px,DaRold:2005mxj,Karch:2006pv}. 
It consists  in properly 
modifying the five dimensional 
classical theory to  resemble  QCD as much as possible.
In this analysis, we used  the so called 
soft-wall (SW) models where  a dilaton field  is introduced to softly 
break 
conformal invariance. 
This model has been successfully applied to investigate non 
perturbative features of hadrons and has been
 extended to study the 
glueball 
spectrum~
\cite{Rinaldi:2017wdn,
Karch:2006pv,Capossoli:2015ywa,Erlich:2005qh,Colangelo:2008us}. 
We 
describe the glueball lattice spectrum 
by means of the AdS/QCD correspondence and then we compare it  with the 
spectrum of 
$f_0$'s, experimentally 
determined \cite{Patrignani:2016xqp}.
Only when the masses of the glueballs and the mesons 
are close, mixing is to be expected~\cite{Vento:2004xx}. However, 
if this mass condition is
 associated  to a  different dynamics, mixing 
will not happen~\cite{Vento:2015yja}. 
Therefore, we are looking for meson and glueball states with similar 
masses but generated by different dynamics.

\section{Scalar glueball and scalar meson spectrum in a bottom-up 
approach}

In the  botton-up approach on the AdS/CFT, one implements  duality in 
nearly 
conformal 
conditions defining QCD on the four dimensional  boundary and 
introducing a bulk 
space which is a slice of $AdS_5$ whose size is  related to 
$\Lambda_{QCD}$ 
~\cite{Brodsky:2003px,DaRold:2005mxj,Erlich:2005qh}. 
The metric of this space can be written as

\vskip -0.5cm
\begin{equation}
ds^2=\frac{R^2}{z^2} (dz^2 + \eta_{\mu \nu} dx^\mu dx^\nu) + 
R^2 d\Omega_5,
\label{metric5}
\end{equation}
where $\eta_{\mu \nu}$ is the Minkowski metric. Within the bottom-up
SW approach, glueballs 
arise from the correspondence with a  field in $AdS_5$ 
~\cite{Colangelo:2008us}.  In our recent work we have 
described the scalar glueball spectrum as that of a graviton in 
$AdS_5$ with a 
modified SW metric~\cite{Rinaldi:2017wdn}.

\subsection{Scalar glueballs and mesons in the soft wall model}
 For a scalar field in the AdS space within the SW model, the 
associated mode function can be obtained  from the equation of 
motion (EoM), see  e.g. Ref. \cite{Capossoli:2015ywa}.

\vskip -0.6cm
\begin{align}
 -\phi_\chi{''}(z)+ \Big[ \kappa^4 z^2+ \frac{15-4 M^2_{5\chi}R^2}{4 
z^2}+2\kappa^2  
\Big]\phi_\chi(z) = \mu_\chi^2 
\phi_\chi(z)~,
\label{scro}
\end{align}
where here, $\chi$  identifies 
  a 
scalar  glueball or a meson field ($\chi=g,m$ respectively), 
$M^2_{5\chi}$  is the $AdS_5$ mass with  $M_{5g}=0$ while 
for the meson  $R^2 M^2_{5m}=-3$. 
In the above relation 
 $\mu_\chi$ is the mass of the $\chi$ field.
The solution of 
the above equation  leads to the a spectrum:
$ \mu^2_{n_\chi} = \Big[4n_{\chi}+C_\chi \Big]\kappa_\chi^2~$,
where  $n_\chi$ represents the mode number and $\kappa_\chi$ the mass 
scale (in principle it could 
depend on the kind of hadron), $C_m=6$ and $C_g=8$.
The corresponding normalised mode functions 
can be written as:

\begin{align}
\label{wf1}
&\Psi_{n_g}(z)=\sqrt{(n_g+1)(n_g+2)/2} \; e^{-\kappa^2 z^2/2} \;  
z^{5/2} \;
_1F_1(-n_g, 3,\kappa^2 z^2)~,
\\
&\Phi_{n_m}(z)=  \sqrt{2(n_m+1)} \; e^{-\kappa z^2/2} \;z^{3/2} \; 
_1F_1(-n_m, 2, \kappa^2 z^2)~.
\end{align}

\subsection{The softwall-graviton model}
\begin{figure}[htb]
\begin{center}
\includegraphics[scale= 0.7]{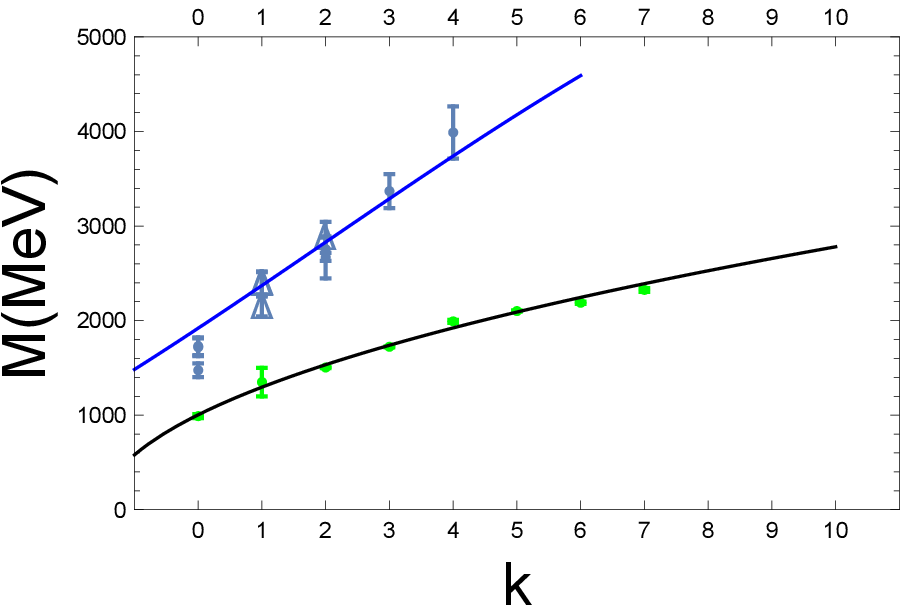}  \hspace{1cm}
\includegraphics[scale= 0.7]{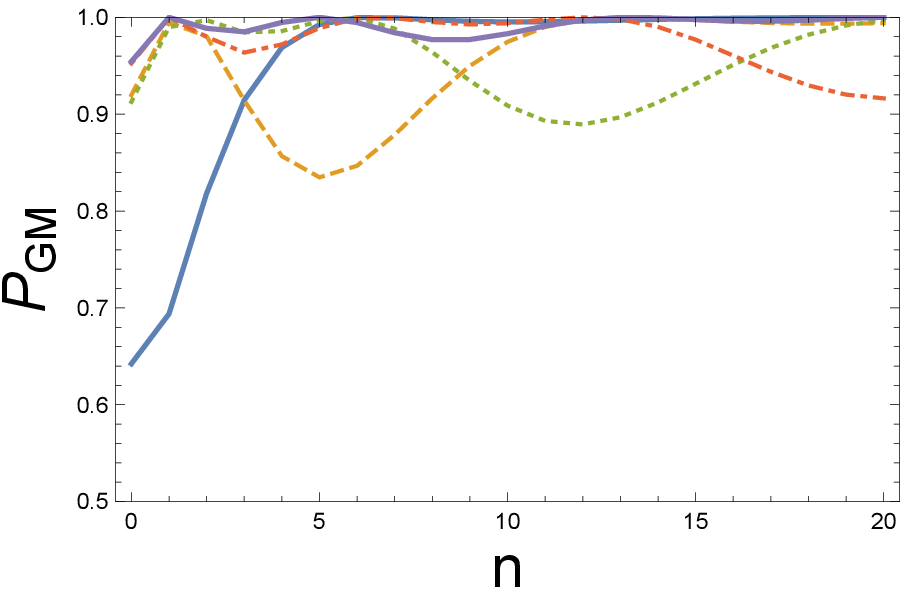} 
 \end{center}
\vskip -0.3cm
\caption{ \footnotesize{ Left panel: fits of glueball spectrum within 
the graviton model (full upper line) and scalar 
meson spectrum (full lower line) obtained within the dilaton SW model 
~\cite{Colangelo:2007pt,Rinaldi:2017wdn}.
 The dark dots represent 
glueball 
spectrum obtained within lattice QCD
~\cite{Morningstar:1999rf,Chen:2005mg,Lucini:2004my}. The light dots 
represent the scalar meson spectrum obtained from experimental data 
~\cite{Patrignani:2016xqp}. Right Panel: We plot the probability of no
 mixing for the glueball with
mode numbers $n_g = 0$ (solid),$1$ (dashed), $2$ (dotted), $3$ 
(dot-dashed), $4$ (solid) as a function of meson mode number $n$. }
}
\label{SpectrumFit}
\end{figure}

In ref.~\cite{Rinaldi:2017wdn}, we discussed the possibility that the 
glueball field is dual to a graviton, to this we generalised the 
background metric:

\vskip -1cm
\begin{align}
\bar g_{MN} = e^{-\alpha^2 z^2/R^2} g_{MN}~,
\end{align}
with $\alpha^2 < 0$ in order to have bound states. The 
EoM is obtained by solving the Einstein equation for the 
above metric~\cite{Rinaldi:2017wdn}:

\vskip -0.8cm

\begin{align}
\label{wf2}
 -\Psi''(t)+\Big[\frac{8 e^{2t^2}}{t^2}-\frac{17}{4 t^2}+14-15 t^2  
\Big]\Psi(t)= \frac{2 \mu^2}{a^2} \Psi(t)~,
\end{align}
where here $t = a z /\sqrt{2}$ and $a = i \alpha$, see 
ref.~\cite{Rinaldi:2017wdn} for details on the spectrum.
In the left panel of Fig. \ref{SpectrumFit} we show our fit of
 the scalar meson and 
glueball spectra obtained within AdS/QCD models.
The fit has been obtained by properly choosing the mass scale 
energy. In the case of mesons (lower line and dots in left panel of Fig. \ref{SpectrumFit}) we 
find a reasonable result for $\kappa \sim 410$ MeV by skipping 
the 
$f_0(500)$ state which
 is not a conventional meson 
state~\cite{Mathieu:2008me,Tanabashi:2018oca}.
For glueballs, the SW model is unable to properly fit 
the lattice spectrum~\cite{Rinaldi:2017wdn}. Within the graviton model, 
the best fit is found for  
$\alpha \sim 370$ 
MeV, see
 upper line of the left panel of 
Fig. \ref{SpectrumFit}.
In this case, the spectrum has an almost linear behaviour 
  with no softening of the slope in the analysed region.
Due to the   difference between the slopes of the glueball 
and 
meson fits for large mode numbers, 
 the related mode  function for a meson will oscillate more than 
that for
 a glueball with similar mass. 
Therefore the large difference in 
mode numbers could be interpreted as a weak mixing between the statess.

\section{Glueball-Meson mixing}
Here
we show the  formalism to
 look for dynamical regions were mixing is not favorable and 
therefore states with mostly gluonic valence structure might exist.
We consider here 
an
holographic light-fron (LF) 
representation of the EoM, in $AdS$ space. The latter can be recast in 
the form 
of a LF
Hamiltonian~\cite{Brodsky:2003px}

\vskip -0.5cm
\begin{equation}
H_{LC} |\Psi_k> = M^2 |\Psi_k>.
\end{equation}
 In this equation both quantities are 
adimensional.
We consider a two dimensional Hilbert space generated by  
a meson 
and a glueball states, \{$|\Psi_m>, |\Phi_g>$\}. 
Mixing occurs when the hamiltonian is not 
diagonal in the subspace. A matrix representation of the hamiltonian 
in 
this subspace is given by

\vskip -0.4cm
\begin{equation}
[H]=  \left( \begin{array}{cc}
m_1 &  \alpha  \\
\alpha & m_2 \end{array} \right) ,
\label{mixing}
\end{equation}
where $\alpha = <\Psi_m|H|\Phi_g>$, $m_1 = <\Psi_m|H|\Psi_m>$ 
and $m_2 = <\Phi_g|H|\Phi_g>$. We are assuming $m_2>m_1$ and for 
simplicity $\alpha$ real and positive. Large $N_C$ QCD analysis shows 
that 
$m_1,m_2 \sim \mathcal{O}(N_C^0)$ and 
$\alpha \sim ~\mathcal{O}(N_C^{-1/2})$~\cite{Vento:2004xx}. 
After diagonalization the eigenstates have a mass
$M_{\pm}= m  \pm \sqrt{ \alpha^2 + 
(\Delta m)^2}$,
where $m=(m_1+m_2)/2$ and $\Delta m = (m_2-m_1)/2$.
The first physical meson, assuming to be the 
lightest one, is given by the eigenvector of $H$, see Ref. 
\cite{Rinaldi:2018yhf}.
In our fit we have fixed the meson spectrum to the experimental values 
and therefore $|\Psi_{phy}>$ represents a physical meson state while we 
have fixed the glueball spectrum to the lattice values, therefore the 
glueball state 
is our initial state $ |\Phi_g>$, thus

\vskip -0.3cm
\begin{equation}
|<\Psi_{phy}|\Phi_g>|^2 = \frac{\alpha^2}{(M_- - m_2)^2}.
\end{equation}
The mixing probability is proportional to the 
overlap  of these two wave functions (w.f.).
We consider Eq. (\ref{wf1}) for the meson w.f. and that from Eq. 
(\ref{wf2}) for the glueball one.
We calculate 
 the  probability for no mixing, i.e., $P_{GM} =1- | \langle 
\Psi_{phy}|\Phi_g\rangle |^2$.
As one can see in the right panel of
 Fig.~\ref{SpectrumFit}, 
  the  mixing should occur when $n_g=2,3,4$ and, following the similar 
mass condition, for meson mode numbers such as $n \sim 10,13,17$.
As one can  see in the right panel of Fig.~\ref{SpectrumFit}
 this condition reduces the  overlap probability for 
mixing dramatically. The outcome of our analysis is 
that the AdS/QCD approach predicts the existence of almost pure 
glueball states in the scalar sector in the mass range above $2$ GeV.

\section{Conclusion}
We have performed a phenomenological analysis of the scalar glueball
and scalar
meson spectrum based on the  AdS/QCD 
correspondence within the SW  and 
softwall-graviton approaches.
Theoretical outcomes have been compared with lattice $QCD$ data in 
the 
case of the 
glueballs and the experimental $f_0$ spectrum of the PDG tables in
 the case of the mesons.    Assuming a light-front 
quantum mechanical description of AdS/QCD correspondence, we have shown 
that the overlap probability of heavy glueballs to heavy mesons is 
small and 
thus one expects little mixing in the high mass sector. Therefore, 
this is the kinematical region to look for almost pure glueball states.

\section*{Acknowledgments}
The author thanks all the orginzers of the conference for the support 
given for 
this talk.
M.R. thanks Vicente Vento for
discussions.  This  work  was  supported in part by the STRONG-2020 
project of the 
European Union's Horizon 2020 research and innovation programme under 
grant agreement No 824093.

\bibliography{iopart-num2}

\end{document}